\documentstyle [12pt]{article}
\title{
\hspace{3.0truein}{\small IFT-485-UNC}\\
\vspace{1.0truein}
{Renormalization Group Flow of Four-Fermi
with Chern-Simon Interaction
}}
\author{Wei Chen\footnotemark[1]\\
          Department of Physics\\
          University of North Carolina\\
          Chapel Hill, NC 27599-3255}
\date{ } 
\footnotetext{$^*$ chen@physics.unc.edu}
\begin{document}
\maketitle
\vspace{0.2truein}

\begin{abstract}
We introduce Chern-Simons interaction into the
three dimensional four-fermi theory, and suggest
a possible line of non-Gaussian infrared
stable fixed points of the four-fermi operator, which is
characterized by the Chern-Simons coupling.
\end{abstract}

\renewcommand{\theequation}{\thesection.\arabic{equation}}
\baselineskip=18.0truept
\newpage
\vskip 0.5truein
\section{Introduction}
\setcounter{equation}{0}
\vspace{3 pt}

Four-fermi theory in $d$ spacetime dimensions ($2<d<4$)
in the low-energy regime is thought
as {\it trivial} as the free theory, since
the only infrared (${\it IR}$)
stable fixed point of four-fermi coupling
is Gaussian. Evidently, this is attributed to
the four-fermi operator being
irrelevant or non-renormalizable.
On the other hand, it was conjectured long ago \cite{FP}\cite{KGW}
that the four-fermi operator could perhaps be relevant if one considered
some non-trivial resummation of infinite numbers of diagrams. And
there have been numerous recent efforts
to explore new phenomena of the theory
in the high-energy regime
\cite{RWP}-\cite{CMS}.
It turns out
that if the four-fermi interaction is {\it attractive} and
if the number of fermion species N is large,
there exists a strong coupling
ultraviolet ({\it UV}) stable fixed point and
the perturbation expansion over the parameter 1/N
is controllable. This {\it UV} fixed point relates to
the dynamical symmetry breaking characterized
by the vacuum expectation $<\psi^\dagger\psi>$
($ \neq 0$ for the
symmetry broken phase). There have been also recent arguments that
near certain strong critical couplings, the four-fermi operator might
be relevant \cite{SW}.
In this Paper, limiting to three dimensions,
we introduce a Chern-Simons interaction, and suggest
a possible line of non-trivial {\it IR}
fixed points for the four-fermi operator, which
is characterized by the Chern-Simons coupling.

Chern-Simons term appears naturally in planar systems.
For instance, the massive fermion current-current
correlation contains an anti-symmetric
part (in the spacetime indices), which
corresponds to an induced Chern-Simons term
in the effective Lagrangian \cite{DJT}.
On the other hand, charged matter fields can be coupled
to a gauge potential who's dynamics is governed
by a Chern-Simons term. Physically, this is to
attach particles with `magnetic' flux tubes.
The composite particles
carrying both charges and flux tubes are
known as anyons \cite{WZ}\cite{W}, and
have been generally thought responsible
to phenomena in the quantum Hall systems
\cite{FHL}-\cite{CFW}.
Numerous efforts have been also made on using the idea of anyons to
understand the high $T_c$ superconductivity \cite{CHWW}\cite{REF},
and other strongly correlating planar systems like the quantum
Heisenberg antiferromagnet \cite{LRE}.

The Chern-Simons coupling,
denoted by a real number $\alpha$,
characterizes the statistics of the matter field, therefore
it is also called the statistical parameter. As $\alpha$
varies from zero to one
[see  (\ref{action}) for the convention], the minimally
coupled matter
changes continuously from the free fermion
to free (hard corn) boson, via anyons
- the very complicated many-body systems.
This is known as the bose-fermi
transmutation \cite{P}-\cite{SSS}.
In relativistic systems, besides statistics transmutation there happens
spin transmutation of matter fields \cite{CI}.
That is, an integer part of $\alpha$ can be reabsorbed
by changing the spin -- the character of the
Poincare representation -- of
matter fields.

Without a reference to the spacetime metric,
and as a top form in three dimensional
(or any odd dimensional) manifold,
Chern-Simons terms are topological ones.
This determines  the Chern-Simons couplings
to be insensitive to changes of the energy scale.
In other words, the beta function of the
Chern-Simons couplings is identically vanishing.
Consequently, the Chern-Simons couplings may serve
well as a controlling parameter in the
perturbation expansion.
It is known 
that the Chern-Simons
coupling characterizes as well the
scaling dimensions of the matter field operators
and composite field operators  \cite{CSW}\cite{CL}.
These scaling dimensions determine the asymptotic
behavior of the anyon systems.
Amusingly, the effect of Chern-Simons couplings on the
scaling dimensions of
matter fields and a class of gauge invariant composite field
operators 
is to drive these operators in the
direction of {\it increased} relevance \cite{CL}.

These observations on the Chern-Simons interaction
are very appealing to the search
for possible {\it IR} fixed points in the four-fermi
theory in three dimensions. Let us couple the fermion minimally
to the Chern-Simons gauge field.
Then as the Chern-Simons coupling $\alpha$ goes up from $0$ to $1$,
the spin-1/2 fermion system turns to be anyons and becomes a
spin-0 (hard core) boson system at $\alpha = 1$. Particularly,
the four-fermi interaction at $\alpha = 0$
turns to be a four-bose interaction at $\alpha=1$.
As is known, while the four-fermi theory has a Gaussian
{\it IR} fixed point, the scalar $\phi^4$ theory in three dimensions
has a Gaussian {\it UV} fixed point and a
non-Gaussian {\it IR} fixed point
\cite{WK}\cite{BGZJ}. Then it is natural to expect,
between the Chern-Simons couplings  $\alpha = 0$ and $1$,
there exists a critical $\alpha_c$ so that as $\alpha$ approaches to
$\alpha_c$,
a transition happens that the system
changes from the fermion-like to boson-like or vise versa.

It is conceivable that this critical $\alpha_c$
has something to do with the critical
scaling dimension of the four-fermi operator. As is known,
the engineering dimension of four-fermi operator in
three spacetime dimensions is four and so the operator is
irrelevant. Now, interacting to the Chern-Simons field, the
operator receives an anomalous dimension, which is a monotonically
decreasing function of $\alpha$ \cite{CL}. Therefore the critical
$\alpha_c$ can be so determined that at which the four-fermi operator
becomes marginally irrelevant.
We shall show in this Paper by the renormalization group approach
that there exists indeed such a critical $\alpha_c$, and when $\alpha >
\alpha_c$ there exists  a non-Gaussian
{\it IR} fixed points of the four-fermi coupling, besides
a Gaussian {\it UV} fixed point.
And when the critical Chern-Simons coupling $\alpha_c$ is
approached from the boson-like side, {\it i.e.} $\alpha (>\alpha_c)\rightarrow
\alpha_c$,
the two fixed points of the four-fermi coupling coincide.
Namely, the Gaussian
fixed point at $\alpha=\alpha_c$
is a multi-fixed point.

The Paper is organized as follows:
Section 2 is devoted to an analysis of the Chern-Simons coupling
as a controlling parameter and an analysis of loop corrections
to the two- and four-point correlation functions.
In Section 3, we solve the renormalization group equation for
the four-point function.
The critical Chern-Simons coupling $\alpha_c$ is determined,
and a picture for the renormalization group flows of the
four-fermi interaction is drawn.
Section 4 is for conclusions and discussions.
In three appendixes,
we present our conventions and calculation details.

\vskip 0.5truein
\section{CS Coupling as a Controlling Parameter}
\setcounter{equation}{0}
\vspace{3 pt}
Let us introduce the topological Chern-Simons term and a
minimal interaction between the Chern-Simons and fermion fields.
The action in Euclidean space reads
\begin{equation}
S = \int d^3x \left(\psi^\dagger\gamma\cdot(\partial-iA)\psi
                  + \frac{G}{2}(\psi^\dagger\psi)^2
+ \frac{i}{4\pi\alpha}
\epsilon^{\mu\nu\lambda}A_\mu\partial_\nu A_\lambda\right)\;.
\label{action}
\end{equation}
Here, for simplicity, we have omitted a fermion mass term\footnotemark.
\footnotetext{
Fermion mass is certainly an interesting issue in the four-fermi
model. Unlike in 4d, there is no chiral symmetry to rule out
a fermion mass term from the lagrangian in 3d.
Instead, the symmetries a fermion mass term in 3d
breaks are parity (P) and time reversal (T). However, since a Chern-Simons
term breaks P and T, the CS coupling will induce a fermion mass term
even the bare fermions are massless, besides the possible dynamical
(mass) generation.
Physically, omitting the fermion mass amounts to choosing
the $M_r=0$ slice on the phase diagram, with $M_r$ denoting the
renormalized fermion mass, as $M_r=0$ is always a fixed point.
This can be realized by introducing
infinite and finite local counter terms to
cancel any induced mass terms.}
Namely, we concentrate on two parameters only, the Chern-Simons
coefficient $\alpha$ and the four-fermi coupling $G$.
With the gauge coupling, this theory is invariant under the
gauge transformations
\begin{eqnarray}
A_\mu(x) &\rightarrow& A_\mu(x) + \partial_\mu\epsilon(x)\;,\\
\psi(x) &\rightarrow& e^{i\epsilon(x)}\psi(x)\;.
\end{eqnarray}
Without the Maxwell term, the Chern-Simons gauge
field carries {\it no} dynamical degree of freedom.
To see this, we vary
(\ref{action}) over $A_\mu$ field, and obtain the equation
of motion for the Chern-Simons field:
\begin{equation}
\frac{1}{2}\epsilon_{\mu\nu\lambda}\partial_\nu A_\lambda
= 2\pi\alpha j_\mu\;,
\label{eqmot}
\end{equation}
where the conserved matter current $j_\mu = i\psi^\dagger\psi$.
Now it is clear that without their own time evolution,
the `electric' and `magnetic' fields associated with the Chern-Simons
gauge potential $A_\mu$ are determined completely
by the spatial and `time' components of the matter
current, respectively. (\ref{eqmot}) can be read in
another way: Taking $\mu=0$, one sees that a
`magnetic' flux
$b=\frac{1}{2}\epsilon_{ij}\partial_i A_j$ in unit
strength $2\pi\alpha$
is attached to a particle.
This attachment of fluxes
makes a change to the statistics the fermion matter fields obey.
Namely that, when the positions of two identical particles are exchanged,
their relative phase is no longer
the exponential of
$i\pi$ but $i(\alpha+1)\pi$.
The real, dimensionless Chern-Simons gauge coupling $\alpha$ is thus
called the statistical parameter.
When $\alpha$ varies from zero to one, the matter field runs
from the fermion limit through anyons to the boson limit, according to
the statistics changes.

Formally, the Chern-Simons interaction is renormalizable and its
perturbation expansion may contain logarithmic divergence. However,
the topological nature of the Chern-Simons term allows only trivial,
finite renormalization of the Chern-Simons term itself.
It has been proven, based on a perturbation expansion \cite{CH}-\cite{C},
a non-renormalization theorem that
says, there is no radiative correction to the (Abelian)
Chern-Simons term  from
massive matters  beyond one loop, and there
is only finite correction to it from massless matters
at even loops (two loops and beyond). This implies
the Chern-Simons coupling is insensitive to changes of energy scale
at all. Consequently, the beta function of the Chern-Simons coupling
identically vanishes
\begin{equation}
 \beta(\alpha) \equiv 0\;.
\end{equation}
Then, it is plausible
to see that the Chern-Simons coupling serves
as a controlling parameter in the
renormalization of
the four-fermi interaction (at least to the
order we are interested in).

On the other hand, the four-fermi coupling constant $G$ has
dimension {\it one} in the unit of length (the inverse of mass).
Introduce a large momentum cutoff,
one can define a dimensionless four-fermi coupling constant $g$ so that
\begin{equation}
G=g/\Lambda\;.
\end{equation}
Accordingly, it is convenient to use the
naive large momentum cutoff $\Lambda$
as a regulator. This regularization procedure
is of course neither gauge nor Lorentz invariant.
We shall make up these in the renormalization.
Considering $\Lambda$ is to sent to infinity eventually
to come back to the continuous theory,
we shall introduce counterterms to cancel
the terms which are
$\Lambda$ dependent  to recover the symmetries.

The zero point renormalization will be  chosen for convenience.
Namely, the renormalized four-fermi
coupling constant is defined at the external momenta
$p_1=p_2=p_3=p_4 = 0$. It is assumed that the asymptotic behavior of
the theory and the existence of non-Gaussian fixed points
are independent of the regularization and renormalization
procedure, though these choices affect in general
the values of fixed point couplings.

We shall consider perturbation
theory over the four-fermi coupling $G$, and calculate the
renormalization group functions of it.
To each order in $G$, we perturb the theory
to the order $\alpha^2$, where one can have almost all
non-trivial observations in the perturbation theory.
We calculate {\it one particle irreducible} diagrams only,
which are sufficient for the purpose of renormalization.
 We shall use the Landau gauge for
fixing the $U(1)$ symmetry. As is well known \cite{PR},
this gauge choice in the Chern-Simons theories
is especially good at avoiding the infrared
divergence in Feynman diagrams.
Leaving the calculation details to the appendixes II and III,
we shall outline the analysis and present the main results in the
rest of the section.

Calculate the renormalization constant of $\psi$ field,
$Z_\psi$, first. Since we plan to consider the four-fermi
vertex to order
$G^2$, it is sufficient to have the fermion wave-function
renormalization to order $G^1$. It is easy to check the fermion bubble
of the fermion self-energy (see Fig.~3a)
vanishes as ${\rm Tr}\gamma_\mu=0$. Then at order $G\alpha^0$,
$Z_\psi = 1$.

At order $G^0\alpha$, it appears a term linear in $\Lambda$
in both the fermion and Chern-Simons
self-energies. This requires mass renormalization. We simply
remove them by introducing the mass counterterms to keep the
theory massless.

At order $G^0\alpha^2$, the fermion self-energy
is (see Appendix II)
\begin{equation}
\Sigma(\Lambda,p)=i\not\! p\frac{\alpha^2}{6}\left(ln(\frac{\Lambda}{p})
                   -\frac{4}{3}+\frac{3\pi^2}{32}\right)\;.
\label{abc}
\end{equation}
Here it shows up the logarithmic divergence, which requires
a non-trivial fermion wave-function renormalization. And the value
of the finite constant terms in the bracket is regularization dependent.

At orders $G\alpha$ and $G\alpha^2$, there is no contribution to
the fermion wave-function renormalization, as the fermion
self-energy takes a form
\begin{equation}
G\Lambda(\alpha+a\alpha^2)(\Lambda +b|p|)\;.
\end{equation}
This obviously matters only the fermion mass
($a$ and $b$ are constants), which we
send to zero by fine tuning as discussed above.

In summary, to the order $G^1\alpha^2$, only  (\ref{abc}) is
non-trivial to the fermion wave-function renormalization
constant. The latter  is
\begin{equation}
Z_\psi = 1 +  \frac{\alpha^2}{6}
ln(\frac{\Lambda}{\mu}) +{\cal O}(G^2)\;.
\label{Zpsi}
\end{equation}

Let us turn to consider the loop corrections to the four-fermi
vertex. First, without the Chern-Simons
vertex, the fermion bubbles contribute to the vertex (see Fig.~3b and c) with
\begin{equation}
\Gamma(\Lambda, p_i) = G^2(\frac{2\Lambda}{\pi^2}-\frac{|p_1-p_2|+|p_3-p_4|}
{8})\;.
\end{equation}
Upto order $G^2\alpha$, we obtain (see  Appendix III)
the vertex function at the renormalization point
\begin{equation}
\Gamma(\Lambda, p=0) = -G(1-i\pi\alpha n\cdot\gamma)
+G^2(\frac{2\Lambda}{\pi^2})(1-i\pi\alpha n\cdot\gamma)\;,
\label{gama21}
\end{equation}
where $n\cdot\gamma$ is a unit Dirac vector, $n^2 = 1$.
(\ref{gama21}) carries three messages: 1) there is an induced
non-local four-fermi vertex related to $p\cdot \gamma/|p|$;
2) the new vertex has the
same coupling constant as that the original four-fermi
vertex has, and so they have the same coupling constant
renormalization.
and (3) at the order $\alpha^1$ there is no change comparing to
the fixed point structure of the
four-fermi theory without Chern-Simons interaction (see Appendix I).

The order $G\alpha^2$ is non-trivial in the sense of renormalization as
the logarithmic divergence appears. Here the four-fermi vertex is:
\begin{equation}
\Gamma^{(1,2)}(\Lambda,p) =
- \frac{5}{2}G\alpha^2[ ln(\frac{\Lambda}{p}) + finite]\;.
\label{Gama}
\end{equation}
The {\it finite} part in (\ref{Gama})
is obviously regularization and renormalization dependent.

At order $G^2\alpha^2$, the vertex function at the renormalization point
takes the form
\begin{equation}
\Gamma^{(2,2)}(\Lambda,p=0)=G^2\alpha^2\Lambda(b_1
+b_2ln(\frac{\Lambda}{\mu}))\;.
\end{equation}
Formally, there appears here the logarithmic divergence too.
However, a diagram-by-diagram investigation shows that
{\it all} the diagrams that contribute to the log term $b_2$
contain, as a sub-diagram, one of the diagrams with
logarithmic divergence at the previous order $G\alpha^2$. Namely,
at order $G^2\alpha^2$, the four-fermi vertex  has
no {\it primary} logarithmic divergence. If counterterms are introduced
to remove divergence order by order,
no logarithmic divergence will be seen at order
$G^2\alpha^2$. Therefore, no contribution comes from this order  to
the vertex renormalization. Again, the finite correction $b_1$
is regularization and renormalization dependent.

\vskip 0.5truein
\section{RG Flows of Four-fermi Interaction}
\setcounter{equation}{0}
\vspace{3 pt}
Now, we are ready to calculate the beta function of
the four-fermi coupling. The renormalized
 coupling at scale $\mu$ is defined as
\begin{eqnarray}
-G_r
 &=& Z_\psi^2(\Lambda)\Gamma(\Lambda,p_i=0)\;\\
&=&-G\left(1+a_1\alpha^2+a_2\alpha^2 ln(\frac{\Lambda}{\mu})\right)
   +G^2\Lambda
\left(b_0+b_1\alpha^2+b_2\alpha^2 ln(\frac{\Lambda}{\mu})\right)\;,
\label{Gr}
\end{eqnarray}
where, from (\ref{Zpsi}) and (\ref{Gama}),
\begin{equation}
a_2 = \frac{17}{6}\;;
\label{a2}
\end{equation}
$b_0=2/\pi^2$, as given in section 2; the other dimensionless
constant $a_1, b_1$ and $b_2$ can be calculated from the Feynman diagrams.
All these renormalization constants
{\it except} $a_2$, however, are dependent of regularization
and renormalization schemes, and thus are not very physically meaningful.
Fortunately, as we shall see soon,
these scheme dependent constants determine only
the values of the fixed point couplings,
which are {\it unnecessarily} universal.

Using the renormalization group equation on (\ref{Gr}), we obtain
\begin{eqnarray}
& &-a_2G\alpha^2+
\left(b_0+b_1\alpha^2
+b_2\alpha^2(1+ln(\frac{\Lambda}{\mu}))\right)G^2\Lambda\nonumber\\
&=&\left(1+a_1\alpha^2+a_2\alpha^2ln(\frac{\Lambda}{\mu})
-2G\Lambda(b_0+b_1\alpha^2
+b_2\alpha^2ln(\frac{\Lambda}{\mu}))\right)\beta(G)\;,
\label{RGq}
\end{eqnarray}
where the beta function of the coupling $G$ is defined as
\begin{equation}
\beta(G) \equiv \Lambda\frac{\partial G(\Lambda)}{\partial \Lambda}\;.
\end{equation}
The solution of the RG equation (\ref{RGq}) to the order $G^2\alpha^2$ is
\begin{equation}
\beta(G) = -b_0G^2\Lambda+\left(-a_2G+(b_1-b_0A_1)G^2\Lambda\right)\alpha^2
+(b_2-a_2b_0)G^2\Lambda\alpha^2ln(\frac{\Lambda}{\mu})\;.
\label{bG}
\end{equation}
As the beta function shouldn't depend on the reference scale $\mu$,
the last term in the right hand side of (\ref{bG}) must vanish.
This gives a consistent condition
to the coefficients of the logarithmic terms at the order $G$ and $G^2$:
\begin{equation}
b_2=a_2b_0\;.
\label{relation}
\end{equation}
The condition is verified in the perturbative calculation by
seeing that there is no primary logarithmic divergence in the four-fermi vertex
at order $G^2$ (see the previous section and Appendix III).

We may also define a beta function for the dimensionless coupling
$g=G\Lambda$:
\begin{eqnarray}
\beta(g) &\equiv & \Lambda\frac{\partial g(\Lambda)}{\partial \Lambda}\;,\\
&=&(1-a_2\alpha^2)g+\left(b_0 + (b_1-b_0a_1-b_0a_2)\alpha^2\right)g^2\;.
\label{beta1}
\end{eqnarray}
{}From (\ref{a2}) and (\ref{beta1}), we obtain the critical Chern-Simons
coupling
\begin{equation}
\alpha_c^2=\frac{1}{a_2}=\frac{6}{17}\;,
\end{equation}
at which the four-fermi operator has scaling dimension three and thus
becomes marginal. Please notice the critical coupling $\alpha_c$ is
gauge choice independent as it is calculated from the anomalous dimensions.

Next, we discuss the nature of fixed points
$g^*$, which by definition satisfy
\begin{equation}
\beta(g^*) = 0\;.
\end{equation}
Obviously, $g=0$ is a fixed point as always.
When $\alpha < \alpha_c$, the slop of the beta function
$\beta^\prime(0)$ is {\it positive},
and therefore the Gaussian fixed point is IR stable.
Meanwhile, the four-fermi operator has a scaling dimension
$(4-\frac{17}{6}\alpha^2) > 3$,
and so is irrelevant. Therefore the matters in
the region  $\alpha < \alpha_c$ are the fermion-like.

When $\alpha > \alpha_c$, matters fall in the boson-like phase.
The slop of the beta function at $g=0$ is {\it negative} and thus
the Gaussian fixed point is IR {\it unstable}. This implies
the renormalization group trajectory must flow to
some {\it IR} stable fixed point as the energy scale is decreasing.
Setting (\ref{beta1}) zero,
a non-Gaussian fixed point can be solved readily, which is
a function of the Chern-Simons coupling $\alpha$,
\begin{equation}
g^* = -\frac{1}{b^2_0}(b_0+b_0a_1\alpha^2-b_1\alpha^2)\;.
\end{equation}
As the slop of the beta function at the non-Gaussian fixed
point is {\it positive}
\begin{equation}
\beta^\prime(g^*) = a_2\alpha^2-1 > 0,~~~
{\rm when~}\alpha > \alpha_c\;,
\label{slop}
\end{equation}
the non-Gaussian fixed point $g^*$ is of the {\it IR} stable.
By adjusting
$\alpha$ $(> \alpha_c)$, $g^*$ can be
as close to $g=0$ as possible, so the perturbation expansion
in $g$ or $G$ is well-defined and controllable.

The renormalization group flows of the four-fermi coupling
are schematically depicted in Fig.~1.

\unitlength=1.00mm
\linethickness{0.4pt}
\thicklines
\begin{picture}(120.0,50.0)
%
%
\put(45.0,15.00){\vector(-1,-1){4}}
\put(49.0,20.00){\vector(-2,-3){3}}
\put(51.75,27.0){\vector(-1,-3){2}}
\put(53.0,29.00){\makebox(0,0)[l]{$\alpha < \alpha_c$}}
\put(46.50,10.50){\vector(-1,0){5.5}}
\put(53.250,12.700){\vector(-3,-1){5.5}}
\put(59.50,16.250){\vector(-3,-2){5}}
\put(65.10,21.60){\vector(-1,-1){4.5}}
\put(67.0,23.0){\makebox(0,0)[l]{$\alpha = \alpha_c$}}
\put(71.0,16.0){\makebox(0,0)[l]{$\alpha > \alpha_c$}}
\put(34.00,10.0){\line(1,0){70.00}}
\put(104.0,10.0){\vector(1,0){5}}
\put(110.0,10.00){\makebox(0,0)[l]{$g(\alpha)$}}
\put(40.00,0.0){\line(0,1){30.00}}
\put(40.0,30.0){\vector(0,1){5}}
\put(28.0,33.00){\makebox(0,0)[l]{$\beta(g)$}}
\put(40.0,10.00){\makebox(0,0)[c]{$*$}}
\put(63.5,10.00){\makebox(0,0)[c]{$*$}}
\put(41.,9.0){\vector(1,-1){4}}
\put(46.5,4.50){\vector(3,-1){5}}
\put(52.5,3.0){\vector(3,1){5}}
\put(58.5,5.0){\vector(1,1){4}}
\put(68.70,15.0){\vector(-1,-1){4}}
\end{picture}
\begin{description}
\item[Fig. 1]
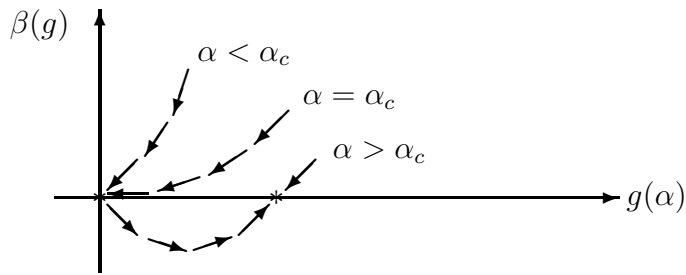
\ \ \ \ Renormalization group flows for the Four-fermi
coupling. Arrows point toward the infrared.
The critical Chern-Simons coupling $\alpha_c$ divides the
anyon matters into two types: the fermion-like
($\alpha < \alpha_c$)
 and the boson-like ($\alpha > \alpha_c$).
\end{description}
\vspace{.5cm}

As is known, the slops of beta function
at fixed points are the anomalous dimensions
of the four-fermi operator, and these are
the physical quantities that determine the
asymptotical behavior of the system.
Their determinations solely by only
the regularization and renormalization {\it independent}
quantity  $a_2$ reflects the consistency of the quantization
procedure used here.

\vskip 0.5truein
\section{Discussions}
\setcounter{equation}{0}
\vspace{3 pt}

We have suggested the exitstence
of a critical Chern-Simons coupling
 $\alpha_c$,  by using perturbation expansion
and renormalization group method.
For a given $\alpha > \alpha_c$, the four-fermi
operator becomes relevant in the low-energy limit,
and there exists a non-Gaussian {\it IR} fixed point
for the four-fermi coupling $g$ (or $G$). This
critical Chern-Simons coupling divides the matters into
two type: the fermion-like and the boson-like, which have
the fixed point structures of the three dimensional fermion
and boson theories, respectively.

However, there is essential difference between
the fermion-like and fermion systems, and between
the boson-like and boson ones. First of all, when
$\alpha \neq 0$ and $1$, the systems describe anyons which
are vary complicated many-body systems, and have not
been understood so well as the fermion and boson systems
have been. And secondly, as is seen,
the Chern-Simons coupling characterizes the universality
classes, and therefore all the anyon systems
with different $\alpha's$ have different critical
exponents, and belong to different critical models.

The fixed point structure of the four-fermi operator
is examined in this work
in the perturbative expansion over $G$. The analysis
here should be reliable, as the non-Gaussian {\it IR}
fixed points $g^*$ can be very small, as long as
$\alpha$ $(>\alpha_c)$ is sufficiently close to $\alpha_c$.
On the other hand, the theory has been perturbed in
the Chern-Simons coupling $\alpha$ as well,
the closer it is to the fermion end $\alpha = 0$,
the more accurate the results are.
As the bose-fermi transmutation strongly suggests the
existence of a critical point $\alpha_c$ between $\alpha = 0$
and $1$, one is tempted to extrapolate the
perturbative results to the not-weak Chern-Simons
couplings such as  $\alpha^2 \sim a_c^2 = 6/17$,
where
qualitatively correct conclusions are drawn,
as suggested in this work.

But one must be very careful with an
extrapolation of large Chern-Simons couplings,
as in the extreme case such as $\alpha \sim 1$, the
perturbative results can not be trusted even qualitatively.
To see this, let's consider three operators
in the fermion Chern-Simons
model: $\psi$, $\psi^\dagger\psi$, and $(\psi^\dagger\psi)^2$.
Their scaling dimensions upto the second order in $\alpha$ are
\begin{equation}
1-\frac{1}{6}\alpha^2, ~~~ 2-\frac{8}{3}\alpha^2,
{}~~{\rm and}~~ 4-\frac{17}{6}\alpha^2\;,
\label{aa}
\end{equation}
respectively \cite{CL}.
Based on the bose-fermi transmutation, on the
other hand, these ``fermion'' operators in the Chern-Simons
fermion theory at $\alpha = 1$ correspond to
the operators in the scalar theory $\phi$, $\phi^*\phi$, and
 $(\phi^*\phi)^2$.  These have scaling dimensions $1/2, 1,$
and $2$, respectively. Taking $\alpha = 1$ in (\ref{aa}),
one sees large discrepancies between the perturbative results and
that suggested by the bose-fermi transmutation. Particularly,
the mass operator would have a negative dimension
in the extrapolation! This provides an
evidence for the inadequacies of the perturbation expansion
beyond its valid regime. The case is unlike
the $\epsilon~ =4-d$ expansion
used in understanding the fixed point structure of the
$\phi^4$ theory in three dimensions \cite{WK}\cite{BGZJ},
where the extrapolation of the controlling
parameter $\epsilon$ from $0$
to $1$ gives reasonable values of exponents.
To improve estimates, certain techniques can be used in perturbation
series when an expansion parameter takes large values,
such as Pad$\acute{{\rm e}}$ approximants which are in fact
used in the $\epsilon$ expansion. For example, such
approximant on the scaling dimension of the mass operator
is $\frac{2}{1+4\alpha^2/3}$, which  gives a positive value of $6/7$
at $\alpha = 1$ (I thank the anonymous referee of the current work
for pointing out the approximants). However, essentially,
non-perturbative treatment to the extreme cases in
the Chern-Simons theory is necessary.

On the other hand, as the Chern-Simons coupling has to be
kept small for a reliable perturbative expansion,
a resolution has been resently
suggested for relativistic Chern-Simons matter theories with
considerable large Chern-Simons coefficients.
That is, one can map one Chern-Simons matter theory into another
Chern-Simons matter theory, in the latter the matter field
has higher spin but weaker Chern-Simons interaction \cite{CI}.

Finally, our study on the renormalization of the four-fermi operator
at lower orders seems to suggest the renormalizability of the
Chern-Simons fermion theories with $\alpha$ not smaller than $\alpha_c$,
however, it needs further careful
considerations for a general proof.

The author is grateful to I. Affleck, L. Dolan, M.P.A. Fisher,
 C. Itoi, A. Kovner, G.W. Semenoff, and Y.-S. Wu
for discussions, and especially
to J.W. Gan for his involvement in the early stage
of the program. The work was supported in part by
the  U.S. DOE under contract No. DE-FG05-85ER-40219.

\newpage
\vskip 0.5truein
\section{Appendix I: Conventions}
\setcounter{equation}{0}
\vspace{3 pt}

For the two-component fermions, we use Hermitean $2\times 2$
Dirac Matrices:
\begin{equation}
\gamma^\mu = \sigma^\mu\; ,
\end{equation}
where $\sigma^\mu$ are Pauli matrices, so that
\begin{eqnarray}
\gamma^\mu\gamma^\nu &=& \delta^{\mu\nu}{\bf 1}
 +i\epsilon^{\mu\nu\lambda}\gamma^\lambda\;,\\
{\bf Tr}{\bf 1} &=& 2\;.
\end{eqnarray}

The Feynman rules, read from (\ref{action}) in the Landau gauge,
are given in Fig.~2.

\unitlength=1.00mm
\linethickness{0.4pt}
\thicklines
\begin{picture}(100.0,90.0)

\put(30.0,75.0){\line(1,0){20.}}
\put(65.00,75.00){\makebox(0,0)[cc]{\ \ \ $=1/(i\not\! p)\;,$}}
\multiput(30.0,60.0)(2.00,0.00){11}{\line(3,0){1.00}}
\put(31.00,58.00){\makebox(0,0)[cc]{$\mu$}}
\put(50.00,58.00){\makebox(0,0)[cc]{$\nu$}}
\put(73.00,60.00){\makebox(0,0)[cc]{\ \ \ \ \ $=
-(2\pi\alpha)\epsilon^{\mu\nu\lambda}p^\lambda/p^2\;,$}}
\multiput(40.,40.0)(0.00,2.00){4}{\line(0,3){1.00}}
\put(40.00,40.00){\circle*{2.00}}
\put(40.00,40.0){\line(-1,-2){3.00}}
\put(40.00,40.0){\line(1,-2){3.00}}
\put(64.00,40.00){\makebox(0,0)[cc]{\ \ $=$\ \ \ $i\gamma_\mu\;,$}}
\put(40.00,15.00){\circle*{2.00}}
\put(40.00,15.0){\line(-1,2){2.00}}
\put(40.00,15.0){\line(-1,-2){2.00}}
\put(40.00,15.0){\line(1,2){2.00}}
\put(40.00,15.0){\line(1,-2){2.00}}
\put(63.00,15.00){\makebox(0,0)[cc]{\ \ \ $=-G\;.$}}

\end{picture}
\begin{description}
\item[Fig. 2]
\ \ \ \ The Feynman rules.
\end{description}
\vspace{.5cm}

A naive large momentum cut-off $\Lambda$ is chosen as a regulator.
This is equivalent to
setting up an energy scale for an effective theory,
and so we may define a dimensionless  (bare)
four-fermi coupling relating to $G$ via
\begin{equation}
g = G\Lambda\;.
\label{g}
\end{equation}

In the rest of this appendix, we consider
the four-fermi theory without the Chern-Simons interaction.
The one loop fermion bubbles are
depicted in Fig.~3.

\unitlength=1.00mm
\linethickness{0.4pt}
\thicklines
\begin{picture}(110.0,42.0)
%
%
\put(25.00,18.00){\circle{40.00}}
\put(25.00,11.00){\circle*{2.00}}
\put(25.00,11.0){\line(-2,-1){5.00}}
\put(25.00,11.0){\line(2,-1){5.00}}
\put(22.0,0.00){\makebox(0,0)[l]{$(a)$}}
%
%
\put(58.00,18.00){\circle*{2.00}}
\put(65.00,18.00){\circle{40.00}}
\put(72.00,18.00){\circle*{2.00}}
\put(54.00,12.00){\makebox(0,0)[cc]{$p_1$}}
\put(54.00,24.00){\makebox(0,0)[cc]{$p_2$}}
\put(77.00,24.00){\makebox(0,0)[cc]{$p_3$}}
\put(77.00,12.00){\makebox(0,0)[cc]{$p_4$}}
\put(58.00,18.0){\line(-1,2){2.00}}
\put(58.00,18.0){\line(-1,-2){2.00}}
\put(72.00,18.0){\line(1,2){2.00}}
\put(72.00,18.0){\line(1,-2){2.00}}
\put(62.0,0.00){\makebox(0,0)[l]{$(b)$
}}
\put(105.00,25.00){\circle*{2.00}}
\put(105.00,18.00){\circle{40.00}}
\put(105.00,11.00){\circle*{2.00}}
\put(101.00,7.00){\makebox(0,0)[cc]{$p_1$}}
\put(101.00,29.00){\makebox(0,0)[cc]{$p_2$}}
\put(110.00,29.00){\makebox(0,0)[cc]{$p_3$}}
\put(110.00,7.00){\makebox(0,0)[cc]{$p_4$}}
\put(105.00,25.0){\line(-1,2){2.00}}
\put(105.00,11.0){\line(-1,-2){2.00}}
\put(105.00,25.0){\line(1,2){2.00}}
\put(105.00,11.0){\line(1,-2){2.00}}
\put(102.0,0.00){\makebox(0,0)[l]{$(c)$}}
\end{picture}
\vspace{0.5cm}
\begin{description}
\item[Fig. 3]
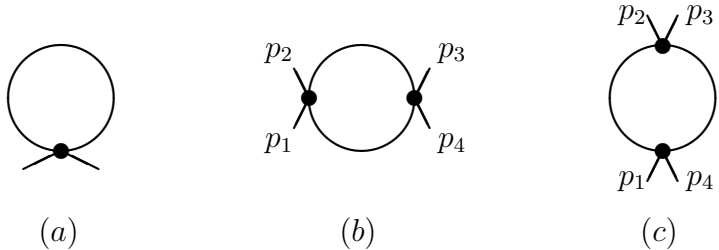
\ \ \ \ One loop diagrams for four-fermi theory. The momenta
$p_1$ and $p_3$ of particles in (b) and (c) are chosen entering
into the loop, and
$p_2$ and $p_4$ of anti-particles outgoing from the loop.
\end{description}
\vspace{.5cm}

\noindent The vanishing of Fig.~3a is obvious
as ${\rm Tr}\gamma_\mu = 0$.
Therefore, the wave-function renormalization
constant to this order is trivial
\begin{equation}
Z_\psi = 1 + {\cal O}(G^2)\;.
\label{Zpsia}
\end{equation}
The correction to the four-fermi vertex from Fig.~1b and 1c is
\begin{equation}
\Gamma(\Lambda, p_i) = G^2(\frac{2\Lambda}{\pi^2}
-\frac{|p_1-p_2|+|p_1-p_4|}{8})\;.
\label{b0}
\end{equation}
We define
the renormalized coupling constant $g_r$ at a
reference energy scale $\mu$ as
\begin{equation}
-\frac{g_r}{\mu} \equiv \Gamma(\Lambda, p_i=0)
= -\frac{1}{\Lambda}(g - \frac{2}{\pi^2}g^2)\;.
\end{equation}
Then the Callan-Symanzik beta function for the dimensionless
coupling $g$ is
\begin{equation}
\beta(g) = g + \frac{2}{\pi^2}g^2 + {\cal O}(g^3) \;.
\label{beta}
\end{equation}
The renormalization group trajectory  
is schematically depicted in Fig.~1
(read $\alpha = 0 < \alpha_c$).
It explicitly shows that near $g=0$, 
the four-fermi coupling
has only a Gaussian IR fixed point,
therefore is {\it trivial}.
This is necessarily consistent with the fact that
the four-fermi operator in three dimensions
is irrelevant or non-renormalizable.

\vskip 0.5truein
\section{Appendix II: Wave-function Renormalization}
\setcounter{equation}{0}
\vspace{3 pt}

We discuss in this Appendix the fermion self-energy, and
calculate the fermion wave-function
renormalization constant to the order $G\alpha^2$.

The fermion and Chern-Simons self-energies at
order $G^0\alpha$ are readily to
calculate

\unitlength=1.00mm
\linethickness{0.4pt}
\thicklines
\begin{picture}(110.00,32.00)
%
%
\put(4.00,12.00){\line(1,0){22.00}}
\multiput(8.100,13.250)(1.00,2.00){4}{\line(3,0){1.00}}
\put(8.10,12.00){\circle*{2.00}}
\multiput(20.900,13.250)(-1.00,2.00){4}{\line(3,0){1.00}}
\put(21.70,12.00){\circle*{2.00}}
\multiput(13.50,20.90)(2.00,0.00){2}{\line(3,0){1.00}}
\put(31.00, 12.00){\makebox(0,0)[l]
{$= i(2\alpha/\pi)\Lambda -i(\pi\alpha/4)|p|\;,$}}
\put(130.0, 12.0){\makebox(0,0)[l]{$(a)$}}
\end{picture}

\unitlength=1.00mm
\linethickness{0.4pt}
\thicklines
\begin{picture}(110.00,28.00)
%
%
\multiput(6.50,18)(-2.00,0.00){3}{\line(3,0){1.00}}
\put(8.00,18.00){\circle*{2.00}}
\put(15.0,18.00){\circle{20.00}}
\put(22.00,18.00){\circle*{2.00}}
\multiput(22.50,18.0)(2.00,0.00){3}{\line(3,0){1.00}}
\put(31.0,18.00){\makebox(0,0)[l]{$
= (1/\pi^2)\Lambda\delta^{\mu\nu}
-(/16)|p|(\delta^{\mu\nu}-p^\mu p^\nu/p^2)\;. $}}
\put(130.0,18.00){\makebox(0,0)[l]{$(b)$}}
\end{picture}
\begin{description}
\item[Fig. 4]
\ \ \ \ (a) Fermion and (b) Chern-Simons self-energies at order
${\cal O} (\alpha)$. There is no
correction at this order to the wave-function renormalization
constants.
\end{description}
\vspace{.5cm}
Fig.~4 carries two messages: The first, there is a need
of mass renormalization for the fermion and
Chern-Simons fields because of the linear terms in
$\Lambda$ in 4a and 4b. We simply get rid of them by
fine tuning with mass counterterms
for both the fermion and Chern-Simons
fields so that they are kept massless. And
the second, there is no contribution
to the wave-function renormalizations at this order,
but it generates finite non-local terms to the effective theory,
due to the second terms in 4a and 4b.

The fermion self-energy diagrams at order $G^0\alpha^2$ are given
in Fig.~5.

\unitlength=1.00mm
\linethickness{0.4pt}
\thicklines
\begin{picture}(110.00,40.00)
%
%
\put(4.00,12.50){\line(1,0){34.00}}
\multiput(9.00,14.00)(1.00,2.00){4}{\line(3,0){1.00}}
\put(9.00,12.50){\circle*{2.00}}
\multiput(32.00,14.00)(-1.00,2.00){4}{\line(3,0){1.00}}
\put(33.00,12.50){\circle*{2.00}}
\multiput(14.00,21.00)(-2.00,0.00){1}{\line(3,0){1.00}}
\put(16.00,21.0){\circle*{2.00}}
\multiput(27.00,21.00)(2.00,0.00){1}{\line(3,0){1.00}}
\put(26.00,21.0){\circle*{2.00}}
\put(21.0,21.00){\circle{10.00}}
\put(20.00,2.00){\makebox(0,0)[cc]{$(a)$}}
\put(53.00,18.0){\line(1,0){35.00}}
\multiput(59.00,19.50)(1.00,2.00){3}{\line(3,0){1.00}}
\put(59.00,18.0){\circle*{2.00}}
\multiput(81.50,19.50)(-1.00,2.00){3}{\line(3,0){1.00}}
\put(82.50,18.0){\circle*{2.00}}
\multiput(63.00,25.00)(2.00,0.00){8}{\line(3,0){1.00}}
\multiput(64.750,16.50)(1.00,-2.00){3}{\line(3,0){1.00}}
\put(64.750,18.0){\circle*{2.00}}
\multiput(75.20,16.50)(-1.00,-2.00){3}{\line(3,0){1.00}}
\put(76.70,18.0){\circle*{2.00}}
\multiput(69.00,11.00)(2.00,0.00){2}{\line(3,0){1.00}}
\put(70.00,2.00){\makebox(0,0)[cc]{$(b)$}}
\put(103.00,18.0){\line(1,0){35.00}}
\multiput(109.00,19.500)(1.00,2.00){3}{\line(3,0){1.00}}
\put(109.0,18.0){\circle*{2.00}}
\multiput(123.50,19.500)(-1.00,2.00){3}{\line(3,0){1.00}}
\put(124.50,18.0){\circle*{2.00}}
\multiput(113.00,25.00)(2.00,0.00){4}{\line(3,0){1.00}}
\multiput(116.750,16.50)(1.00,-2.00){3}{\line(3,0){1.00}}
\put(116.750,18.0){\circle*{2.00}}
\multiput(131.0,16.50)(-1.00,-2.00){3}{\line(3,0){1.00}}
\put(132.0,18.0){\circle*{2.00}}
\multiput(121.00,11.00)(2.00,0.00){4}{\line(3,0){1.00}}
\put(120.00,2.00){\makebox(0,0)[cc]{$(c)$}}
\end{picture}
\vspace{0.4cm}
\begin{description}
\item[Fig. 5]
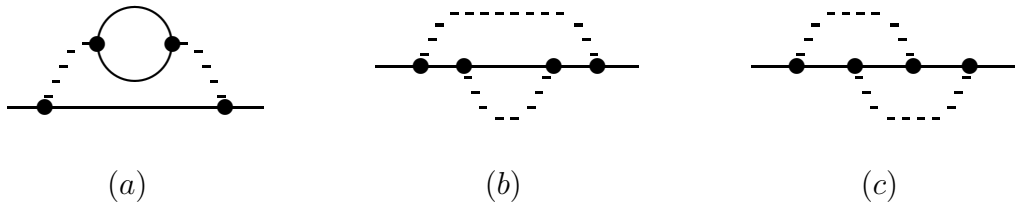
\ \ \ \ The fermion self-energy at ${\cal O}(\alpha^2)$.
Logarithmic divergence appears here.
\end{description}
\vspace{0.5cm}

\noindent Calculating the diagrams, after one of the two
loop integrations is done, we are left with
\begin{eqnarray}
(f.5a)&=&-\frac{i(\pi\alpha)^2}{2} \int^\Lambda\frac{d^3k}{(2\pi)^3}
         \frac{\not\! k (k+p)\cdot k}{(k+p)^2k^3}\;,  \\
(f.5b)&=&-i(\pi\alpha)^2\int^\Lambda\frac{d^3k}{(2\pi)^3}
         \frac{\not\! k}{|k+p|k^2}\;,\\
(f.5c)&=&-\frac{i(\pi\alpha)^2}{2} \frac{\not\! p}{p^2}
          \int^\Lambda\frac{d^3k}{(2\pi)^3}
[\frac{(k+p)\cdot p}{(k+p)^2k}\nonumber\\
    & & ~~~ - \frac{1}{2}\frac{k}{(k+p)^2}
              +\frac{1}{2}\frac{(k-p)\cdot k}{|k+p|k^2} - \frac{p}{|k+p|k}
              +\frac{p(k+\frac{1}{2}p)\cdot k}{(k+p)^2k^2}]\;.
\end{eqnarray}
Completing the integral over $k$,
we obtain the fermion self-energy to this order
\begin{equation}
\Sigma(\Lambda,p)=i\not\! p\frac{\alpha^2}{6}\left(ln(\frac{\Lambda}{p})
                   -\frac{4}{3}+\frac{3\pi^2}{32}\right)\;.
\end{equation}
Here it shows up first the logarithmic divergence, which
obviously requires
a non-trivial fermion wave-function renormalization.

To continue the discussion with the fermion self-energy to
higher orders,
let us consider first these three vertex corrections
that have a fermion loop.
The relevant diagrams are given in Fig.~6. Fig.~6a vanishes
due to its tensor structure: the trace over the Dirac matrices
in the fermion loop gives
$Tr(\gamma_\mu\gamma_\nu\gamma_\lambda)
=2\epsilon_{\mu\nu\lambda}$, while
its contraction with $p_\nu p_\lambda$ from the
loop integration is zero.
It is not difficult to check that Fig.~6b and 6c
are proportional to $p_\mu$, the momentum carried by the
external Chern-Simons line.

\unitlength=1.00mm
\linethickness{0.4pt}
\thicklines
\begin{picture}(110.0,45.0)
%
%
\put(25.0,25.0){\circle*{2.00}}
\put(25.00,18.00){\circle{40.00}}
\put(25.00,11.00){\circle*{2.00}}
\put(25.00,11.0){\line(-2,-1){5.00}}
\put(25.00,11.0){\line(2,-1){5.00}}
\multiput(25.00,26.0)(0.00,2.00){3}{\line(0,3){1.00}}
\put(28.00,29.00){\makebox(0,0)[cc]{$p$}}
\put(22.00,29.00){\makebox(0,0)[cc]{$\mu$}}
\put(35.0,18.00){\makebox(0,0)[l]{$= 0\;;$}}
\put(22.0,0.00){\makebox(0,0)[l]{$(a)$}}
\put(70.0,25.0){\circle*{2.00}}
\multiput(70.00,26.0)(0.00,2.00){3}{\line(0,3){1.00}}
\put(73.00,29.00){\makebox(0,0)[cc]{$p$}}
\put(67.00,29.00){\makebox(0,0)[cc]{$\mu$}}
\put(63.00,18.00){\circle*{2.00}}
\put(77.00,18.00){\circle*{2.00}}
\put(70.00,18.00){\circle{40.00}}
\put(70.00,11.00){\circle*{2.00}}
\put(70.00,11.0){\line(-2,-1){5.00}}
\put(70.00,11.0){\line(2,-1){5.00}}
\multiput(63.00,18.00)(2.00,0.00){7}{\line(3,0){1.00}}
\put(85.0,18.00){\makebox(0,0)[l]{$\sim$}}
\put(68.0,0.00){\makebox(0,0)[l]{$(b)$}}
\put(105.0,25.0){\circle*{2.00}}
\multiput(105.00,26.0)(0.00,2.00){3}{\line(0,3){1.00}}
\put(108.00,29.00){\makebox(0,0)[cc]{$p$}}
\put(102.00,29.00){\makebox(0,0)[cc]{$\mu$}}
\put(110.00,13.50){\circle*{2.00}}
\put(110.00,22.500){\circle*{2.00}}
\put(105.00,18.00){\circle{40.00}}
\put(105.00,11.00){\circle*{2.00}}
\put(105.00,11.0){\line(-2,-1){5.00}}
\put(105.00,11.0){\line(2,-1){5.00}}
\multiput(110.0,14.00)(0.00,2.00){4}{\line(0,3){1.00}}
\put(120.0,18.00){\makebox(0,0)[l]{$\sim p_\mu$}}
\put(102.0,0.00){\makebox(0,0)[l]{$(c)$}}
\end{picture}
\vspace{0.5cm}
\begin{description}
\item[Fig. 6]
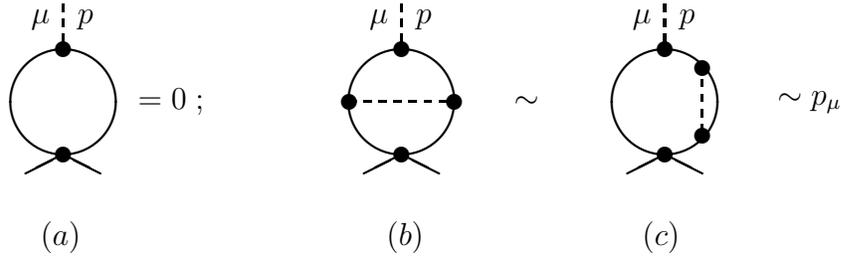
\ \ \ \ The three vertex at ${\cal O}(G)$ and at ${\cal O}(G\alpha)$.
(a) vanishes; (b) and (c) are proportional to $p_\mu$.
\end{description}
\vspace{0.5cm}

The diagrams for
the fermion self-energy at order $G\alpha$ can be obtained by
inserting one Chern-Simons
propagator as an internal line into the fermion self-energy diagram
Fig.~3a in all possible ways.
These are given in Fig.~7.  At this order,
there is no correction to the fermion wave-function renormalization:
Fig.~7a vanishes, because Fig.~3a is zero;
Fig.~7b vanishes as one of its subdiagrams,
shown in Fig.~6a, is zero; And finally,
Fig.~7c is proportional to $\Lambda$ and so contributes
to the fermion mass, which, by fine tuning, is sent to
zero.

\unitlength=1.00mm
\linethickness{0.4pt}
\thicklines
\begin{picture}(110.0,38.0)
%
%
\put(18.00,7.50){\circle*{2.00}}
\put(32.00,7.50){\circle*{2.00}}
\put(25.00,18.00){\circle{40.00}}
\put(25.00,11.00){\circle*{2.00}}
\put(25.00,11.0){\line(-2,-1){10.00}}
\put(25.00,11.0){\line(2,-1){10.00}}
\multiput(17.75,7.50)(2.00,0.00){7}{\line(3,0){1.00}}
\put(22.0,0.00){\makebox(0,0)[l]{$(a)$}}
\put(98.00,18.00){\circle*{2.00}}
\put(112.00,18.00){\circle*{2.00}}
\put(105.00,18.00){\circle{40.00}}
\put(105.00,11.00){\circle*{2.00}}
\put(105.00,11.0){\line(-2,-1){5.00}}
\put(105.00,11.0){\line(2,-1){5.00}}
\multiput(97.75,18.00)(2.00,0.00){7}{\line(3,0){1.00}}
\put(103.0,0.00){\makebox(0,0)[l]{$(c)$}}
\put(72.00,7.75){\circle*{2.00}}
\put(72.00,17.00){\circle*{2.00}}
\put(65.00,18.00){\circle{40.00}}
\put(65.00,11.00){\circle*{2.00}}
\put(65.00,11.0){\line(-2,-1){5.00}}
\put(65.00,11.0){\line(2,-1){11.00}}
\multiput(72.0,8.50)(0.00,2.00){4}{\line(0,3){1.00}}
\put(62.0,0.00){\makebox(0,0)[l]{$(b)$}}
\end{picture}
\vspace{0.5cm}
\begin{description}
\item[Fig. 7]
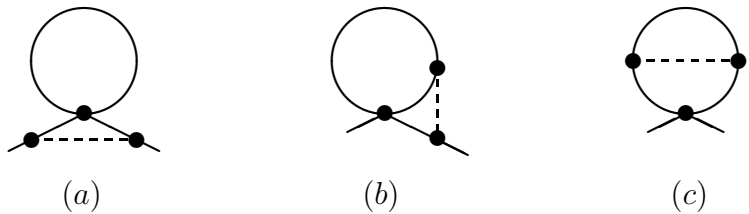
\ \ \ \ The fermion self-energy at ${\cal O}(G\alpha)$.
There is no correction to the wave-function renormalization
at this order.
\end{description}
\vspace{0.5cm}

Finally, at the order  $G\alpha^2$
there is no contribution to the fermion wave-function
renormalization either. This is because,
besides many vanishing diagrams
such as those with any of Fig.~3a, Fig.~6a, 6b and 6c,
and Fig.~7a, 7b and 7c as a sub-diagram,
all the non-vanishing diagrams are of form
\begin{equation}
G\alpha^2\Lambda(a\Lambda +b|p|)\;,
\end{equation}
which matters only the fermion mass
renormalization with a and b finite constants,
as discussed above.
A typical diagram of this type is drawn in Fig.~8.

\unitlength=1.00mm
\linethickness{0.4pt}
\thicklines
\begin{picture}(110.0,38.0)
%
%
\put(72.00,7.75){\circle*{2.00}}
\put(72.00,17.00){\circle*{2.00}}
\put(58.00,7.75){\circle*{2.00}}
\put(58.00,17.00){\circle*{2.00}}
\put(65.00,18.00){\circle{40.00}}
\put(65.00,11.00){\circle*{2.00}}
\put(65.00,11.0){\line(-2,-1){11.00}}
\put(65.00,11.0){\line(2,-1){11.00}}
\multiput(72.0,8.50)(0.00,2.00){4}{\line(0,3){1.00}}
\multiput(58.0,8.50)(0.00,2.00){4}{\line(0,3){1.00}}
\end{picture}
\begin{description}
\item[Fig. 8]
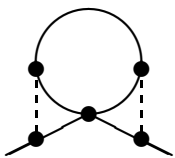
\ \ \ \ A typical non-vanishing diagram of
fermion self-energy at ${\cal O}(G\alpha^2)$.
It takes a form $\Lambda(a\Lambda+b|p|)$.
\end{description}
\vspace{0.5cm}

In summary, the fermion wave-function renormalization
constant is
\begin{equation}
Z_\psi = 1 +  \frac{\alpha^2}{6}
ln(\frac{\Lambda}{\mu}) +{\cal O}(G^2)\;.
\end{equation}

\vskip 0.5truein
\section{Appendix III: Four-Fermion Vertex Function}
\setcounter{equation}{0}
\vspace{3 pt}

In this Appendix, we consider the four-fermi vertex
$\Gamma(\Lambda, p_1,p_2,p_3,p_4)$ to the order
$G^2\alpha^2$. A representative
diagram at order $G\alpha$ is depicted in Fig.~9.

\unitlength=1.00mm
\linethickness{0.4pt}
\thicklines
\begin{picture}(110.00,38.00)
%
%
\put(66.00,12.00){\line(1,-1){12.00}}
\put(66.00,12.00){\line(-1,1){6.00}}
\put(66.00,12.00){\line(-1,-1){6.00}}
\put(66.00,12.00){\line(1,1){12.00}}
\multiput(73.,6.)(0,2.00){7}{\line(0,3){1.00}}
\put(73.0,5.50){\circle*{2.00}}
\put(73.0,19.00){\circle*{2.00}}
\put(66.0,12.00){\circle*{2.00}}
\put(62.00,4.00){\makebox(0,0)[cc]{$p_1$}}
\put(62.00,20.00){\makebox(0,0)[cc]{$p_2$}}
\put(80.00,22.00){\makebox(0,0)[cc]{$p_3$}}
\put(80.00,2.00){\makebox(0,0)[cc]{$p_4$}}
\end{picture}
\vspace{0.5cm}
\begin{description}
\item[Fig. 9]
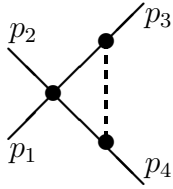
\ \ \ \ Four-fermi vertex at
${\cal O}(G\alpha)$. It generates new finite
term in the effective Lagrangian.
\end{description}
\vspace{0.5cm}

\noindent We calculate Fig.~9, and obtain $(i = 1, 2, 3, 4)$
\begin{equation}
\Gamma^{(1,1)}(\Lambda,p_i)
= \frac{i\pi\alpha G}{8}
\frac{p_3+p_4-|p_3-p_4|}{p_3+p_4+|p_3-p_4|}
(\frac{\not\! p_3}{p_3}+\frac{\not\! p_4}{p_4})
+ {\rm permutation ~in~} p_i \;.
\label{g11}
\end{equation}
Since (\ref{g11}) involves a unit Dirac vector $n\cdot\gamma$
$(n^2 =1)$ which
is not seen in the bare Lagrangian, this means a generation of
a new operator in the effective Lagrangian. An induced new
interaction may probably need a new coupling constant and
so its renormalization.
However, it is not the case in the present model, at least
to the second order.
To see this, let us consider the order $G^2\alpha$.
The diagrams are depicted in Fig.~10.

\unitlength=1.00mm
\linethickness{0.4pt}
\thicklines
\begin{picture}(110.0,88.0)
\multiput(25.0,64.0)(0.00,-2.00){7}{\line(0,3){1.00}}
\put(25.00,65.00){\circle*{2.00}}
\put(25.00,51.00){\circle*{2.00}}
\put(18.00,58.00){\circle*{2.00}}
\put(25.00,58.00){\circle{40.00}}
\put(32.00,58.00){\circle*{2.00}}
\put(18.00,58.0){\line(-1,2){2.00}}
\put(18.00,58.0){\line(-1,-2){2.00}}
\put(32.00,58.0){\line(1,2){2.00}}
\put(32.00,58.0){\line(1,-2){2.00}}
\put(22.0,40.00){\makebox(0,0)[l]{$(a)$}}
\multiput(60.0,62.50)(2.00,0.00){5}{\line(3,0){1.00}}
\put(60.00,62.50){\circle*{2.00}}
\put(70.50,62.50){\circle*{2.00}}
\put(58.00,58.00){\circle*{2.00}}
\put(65.00,58.00){\circle{40.00}}
\put(72.00,58.00){\circle*{2.00}}
\put(58.00,58.0){\line(-1,2){2.00}}
\put(58.00,58.0){\line(-1,-2){2.00}}
\put(72.00,58.0){\line(1,2){2.00}}
\put(72.00,58.0){\line(1,-2){2.00}}
\put(62.0,40.00){\makebox(0,0)[l]{$(b)$}}
\multiput(115.0,68.0)(0.00,-2.00){5}{\line(0,3){1.00}}
\put(115.00,68.00){\circle*{2.00}}
\put(115.00,60.500){\circle*{2.00}}
\put(111.25,52.250){\circle*{2.00}}
\put(114.750,55.750){\circle{10.00}}
\put(105.00,58.00){\line(1,-1){11.00}}
\put(105.00,58.00){\line(-1,1){4.00}}
\put(105.00,58.00){\line(-1,-1){4.00}}
\put(105.00,58.00){\line(1,1){14.00}}
\put(105.0,58.00){\circle*{2.00}}
\put(104.0,40.00){\makebox(0,0)[l]{$(c)$}}
\multiput(54.0,25.0)(0.00,-2.00){8}{\line(0,3){1.00}}
\put(54.00,25.50){\circle*{2.00}}
\put(54.00,10.00){\circle*{2.00}}
\put(58.00,18.00){\circle*{2.00}}
\put(65.00,18.00){\circle{40.00}}
\put(72.00,18.00){\circle*{2.00}}
\put(58.00,18.0){\line(-1,2){6.00}}
\put(58.00,18.0){\line(-1,-2){6.00}}
\put(72.00,18.0){\line(1,2){2.00}}
\put(72.00,18.0){\line(1,-2){2.00}}
\put(62.0,0.00){\makebox(0,0)[l]{$(e)$}}
\multiput(25.0,25.0)(2.00,0.00){5}{\line(3,0){1.00}}
\put(25.00,25.00){\circle*{2.00}}
\put(35.30,25.00){\circle*{2.00}}
\put(18.00,18.00){\circle*{2.00}}
\put(25.00,18.00){\circle{40.00}}
\put(32.00,18.00){\circle*{2.00}}
\put(18.00,18.0){\line(-1,2){2.00}}
\put(18.00,18.0){\line(-1,-2){2.00}}
\put(32.00,18.0){\line(1,2){6.00}}
\put(32.00,18.0){\line(1,-2){2.00}}
\put(22.0,0.00){\makebox(0,0)[l]{$(d)$}}
\multiput(99.0,25.0)(2.00,0.00){8}{\line(3,0){1.00}}
\put(98.50,25.00){\circle*{2.00}}
\put(115.50,25.00){\circle*{2.00}}
\put(102.00,18.00){\circle*{2.00}}
\put(107.00,18.00){\circle{10.00}}
\put(112.00,18.00){\circle*{2.00}}
\put(102.00,18.0){\line(-1,2){6.00}}
\put(102.00,18.0){\line(-1,-2){3.00}}
\put(112.00,18.0){\line(1,2){6.00}}
\put(112.00,18.0){\line(1,-2){3.00}}
\put(104.0,0.00){\makebox(0,0)[l]{$(f)$}}
\end{picture}
\vspace{0.5cm}
\begin{description}
\item[Fig. 10]
\ \ \ \ Four-fermi vertex at ${\cal O}(G^2\alpha)$.
\end{description}
\vspace{.5cm}
Not losing the generality, we set $p_1=p_2=p_3=p_4$.
It is easy to check
\begin{eqnarray}
& &(f.10a)~=~ (f.10b)~=~(f.10c)~=~(f.10d)~=0\;,\\
& &(f.10e)\times 2+(f.10f)\times 2
= -G^2(\frac{2\Lambda}{\pi^2})(i\pi\alpha n\cdot\gamma)\;.
\end{eqnarray}
Therefore, to the order $G^2\alpha$, the vertex function is
\begin{equation}
\Gamma(\Lambda) = -G(1-i\pi\alpha n\cdot\gamma)
+G^2(\frac{2\Lambda}{\pi^2})(1-i\pi\alpha n\cdot\gamma)\;.
\label{gama2}
\end{equation}
This implies that the coupling constant $G$ and its
renormalization is sufficient for both the original
and the induced self-interactions. (\ref{gama2}) also
shows that at the order $\alpha^1$, there is no non-trivial
renormalization to the four-fermi coupling.

Let us come to the next order in $\alpha$.
The Feynman diagrams for the four-fermi vertex
at order $G\alpha^2$ fall into
two categories: the logarithmic divergent and  the finite,
depicted in Fig.~11 and 12, respectively.

\begin{picture}(140.00,120.00)
%
%
\put(30.00,100.00){\circle*{2.00}}
\put(70.00,100.00){\circle*{2.00}}
\put(110.00,100.00){\circle*{2.00}}
\put(30.00,64.00){\makebox(0,0)[cc]{$(a)$}}
\put(70.00,64.00){\makebox(0,0)[cc]{$(b)$}}
\put(110.00,64.00){\makebox(0,0)[cc]{$(c)$}}
\put(30.00,100.0){\line(-1,-2){14.00}}
\put(70.00,100.0){\line(-1,-2){14.00}}
\put(110.00,100.0){\line(-1,-2){14.00}}
\put(30.00,100.0){\line(1,-2){14.00}}
\put(70.00,100.0){\line(1,-2){14.00}}
\put(110.00,100.0){\line(1,-2){14.00}}
\put(30.00,100.0){\line(-1,2){3.00}}
\put(70.00,100.0){\line(-1,2){3.00}}
\put(110.00,100.0){\line(-1,2){3.00}}
\put(30.00,100.0){\line(1,2){3.00}}
\put(70.00,100.0){\line(1,2){3.00}}
\put(110.00,100.0){\line(1,2){3.00}}
\multiput(23.50,86.50)(2.00,0.00){7}{\line(3,0){1.00}}
\put(23.50,86.50){\circle*{2.00}}
\put(36.50,86.50){\circle*{2.00}}
\multiput(18.50,76.50)(2.00,0.00){12}{\line(3,0){1.00}}
\put(18.50,76.50){\circle*{2.00}}
\put(41.50,76.50){\circle*{2.00}}
\multiput(63.50,86.50)(2.00,-1.00){9}{\line(3,0){1.00}}
\put(63.0,86.50){\circle*{2.00}}
\put(80.50,78.50){\circle*{2.00}}
\multiput(75.50,86.50)(-2.00,-1.00){9}{\line(3,0){1.00}}
\put(76.50,86.50){\circle*{2.00}}
\put(59.50,78.50){\circle*{2.00}}
\multiput(98.75,78.00)(2.00,0.00){3}{\line(3,0){1.00}}
\put(99.25,78.0){\circle*{2.00}}
\put(105.25,78.0){\circle*{2.00}}
\multiput(120.50,78.00)(-2.00,0.00){3}{\line(3,0){1.00}}
\put(121.0,78.0){\circle*{2.00}}
\put(115.0,78.0){\circle*{2.00}}
\put(110,78){\circle{10}}
\put(30.00,40.00){\circle*{2.00}}
\put(70.00,40.00){\circle*{2.00}}
\put(110.00,40.00){\circle*{2.00}}
\put(30.00,5.00){\makebox(0,0)[cc]{$(d)$}}
\put(70.00,5.00){\makebox(0,0)[cc]{$(e)$}}
\put(110.00,5.00){\makebox(0,0)[cc]{$(f)$}}
\put(30.00,40.0){\line(-1,-2){14.00}}
\put(70.00,40.0){\line(-1,-2){14.00}}
\put(110.00,40.0){\line(-1,-2){7.00}}
\put(30.00,40.0){\line(1,-2){14.00}}
\put(70.00,40.0){\line(1,-2){14.00}}
\put(110.00,40.0){\line(1,-2){7.00}}
\put(30.00,40.0){\line(-1,2){3.00}}
\put(70.00,40.0){\line(-1,2){3.00}}
\put(110.00,40.0){\line(-1,2){3.00}}
\put(30.00,40.0){\line(1,2){3.00}}
\put(70.00,40.0){\line(1,2){3.00}}
\put(110.00,40.0){\line(1,2){3.00}}
\multiput(33.0,33.5)(2.00,-1.00){4}{\line(3,0){1.0}}
\put(33.0,33.50){\circle*{2.00}}
\multiput(39.50,21.0)(0.00,2.00){5}{\line(3,0){1.00}}
\put(39.50,21.0){\circle*{2.00}}
\multiput(18.50,16.5)(2.00,0.00){12}{\line(3,0){1.00}}
\put(18.50,16.50){\circle*{2.00}}
\put(41.50,16.50){\circle*{2.00}}
\multiput(62.25, 24.25)(2.00,0.00){8}{\line(3,0){1.00}}
\put(62.50,24.50){\circle*{2.00}}
\put(77.50,24.50){\circle*{2.00}}
\multiput(75.0,30.5)(2.00,-1.00){4}{\line(3,0){1.0}}
\put(75.0,30.50){\circle*{2.00}}
\multiput(81.0,18.0)(0.00,2.00){5}{\line(3,0){1.00}}
\put(81.0,18.0){\circle*{2.00}}
\multiput(103.0,25.5)(0.00,-2.00){5}{\line(0,3){1.00}}
\put(103.0,25.50){\circle*{2.00}}
\put(103.0,16.50){\circle*{2.00}}
\multiput(117.0,25.50)(0.00,-2.00){5}{\line(0,3){1.00}}
\put(117.0,25.50){\circle*{2.00}}
\put(117.0,16.50){\circle*{2.00}}
\put(103.0,25.95){\line(1,0){14.}}
\put(103.0,17.00){\line(1,0){14.}}
\put(103.0,17.00){\line(-1,-2){3.0}}
\put(117.0,17.00){\line(1,-2){3.0}}
\end{picture}
\begin{description}
\item[Fig. 11]
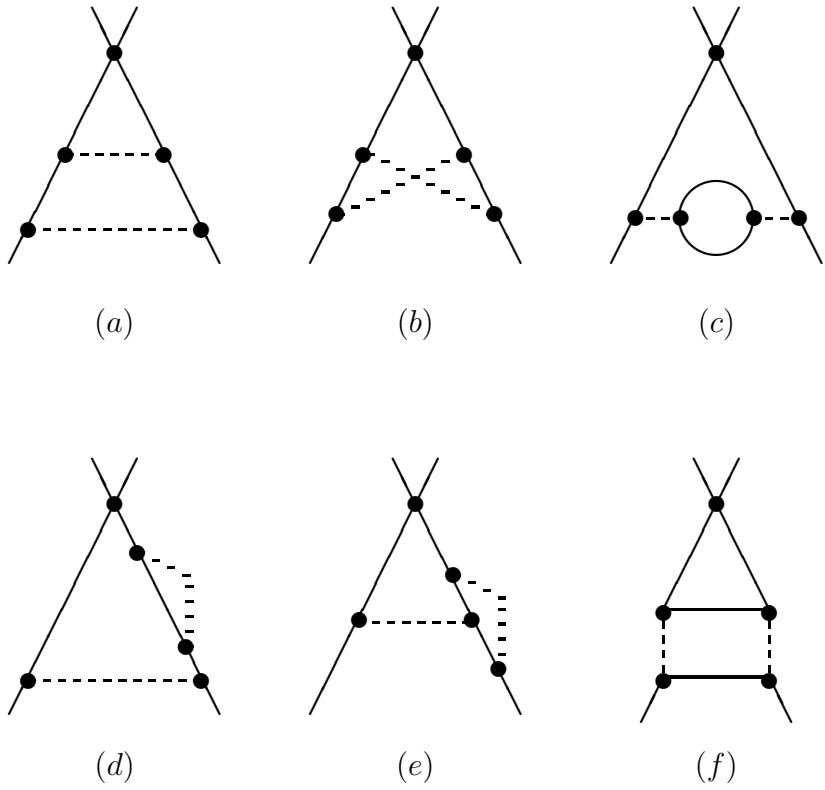
\ \ \ \ Four-fermi vertex at
${\cal O}(G\alpha^2)$: the logarithmic divergent diagrams.
\end{description}
\vspace{0.5cm}

\unitlength=1.00mm
\linethickness{0.4pt}
\thicklines
\begin{picture}(110.00,38.00)
%
%
\put(36.00,12.00){\line(1,-1){12.00}}
\put(36.00,12.00){\line(-1,1){12.00}}
\put(36.00,12.00){\line(-1,-1){12.00}}
\put(36.00,12.00){\line(1,1){12.00}}
\multiput(43.,6.)(0,2.00){7}{\line(0,3){1.00}}
\put(43.0,5.50){\circle*{2.00}}
\put(43.0,19.00){\circle*{2.00}}
\multiput(29.,6.)(0,2.00){7}{\line(0,3){1.00}}
\put(29.0,5.50){\circle*{2.00}}
\put(29.0,19.00){\circle*{2.00}}
\put(36.0,12.00){\circle*{2.00}}
\put(32.00,-5.00){\makebox(0,0)[l]{$(a)$}}
\put(96.00,12.00){\line(1,-1){12.00}}
\put(96.00,12.00){\line(-1,1){12.00}}
\put(96.00,12.00){\line(-1,-1){12.00}}
\put(96.00,12.00){\line(1,1){12.00}}
\multiput(99.,17.)(-2.,0.0){4}{\line(3,0){1.00}}
\put(101.0,17.00){\circle*{2.00}}
\put(91.0,17.00){\circle*{2.00}}
\multiput(104.,5.)(0,2.00){8}{\line(0,3){1.00}}
\put(104.0,4.50){\circle*{2.00}}
\put(104.0,20.00){\circle*{2.00}}
\put(96.0,12.00){\circle*{2.00}}
\put(92.00,-5.00){\makebox(0,0)[l]{$(b)$}}
\end{picture}
\vspace{1.cm}
\begin{description}
\item[Fig. 12]
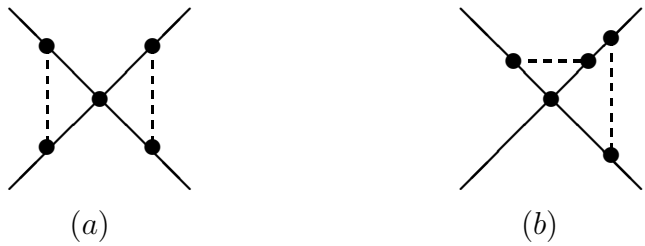
\ \ \ \ Four-fermi vertex at
${\cal O}(G\alpha^2)$: the finite diagrams.
\end{description}
\vspace{0.5cm}

After integrating out one of the two Feynman integrations
of the logarithmic divergent diagrams in Fig. 11, we arrive at
\begin{eqnarray}
(f.11a)+(f.11d)\times 2 &=&-G(\pi\alpha)^2\int^\Lambda\frac{d^3k}{(2\pi)^3}
\frac{1}{(k+p)^2k}\;,\\
(f.11b)+(f.11e)\times 2 &=&-\frac{3}{2}G(\pi\alpha)^2
\int^\Lambda\frac{d^3k}{(2\pi)^3}
\frac{1}{(k+p)^2k}\;,\\
(f.11c)+(f.11f)\times 2 &=&-\frac{5}{2}G(\pi\alpha)^2
\int^\Lambda\frac{d^3k}{(2\pi)^3}
\frac{1}{(k+p)^2k}\;.
\end{eqnarray}
Then, completing the integral over $k$,
we obtain
\begin{equation}
\Gamma^{(1,2)}(\Lambda,p) =
- \frac{5}{2}G\alpha^2[ ln(\frac{\Lambda}{p}) + finite]\;,
\end{equation}
where, the {\it finite} part, from Fig.~11 and 12,
is regularization and renormalization dependent.

At the order $G^2\alpha^2$, most of
the four-fermi vertex Feynman diagrams can be obtained
by inserting one Chern-Simons propagator as an internal line
into all the diagrams in Fig.~10.
As the number of such diagrams
is so huge, we decided to not draw them here
but present the result of our analysis.
These (three-loops) diagrams belong to one of the
three types: (1) vanishing, (2) linear in $\Lambda$, and (3)
proportional to $\Lambda ln(\frac{\Lambda}{\mu})$.
The vertex function at the renormalization point takes the form
\begin{equation}
\Gamma^{(2,2)}(\Lambda,p=0)=G^2\alpha^2\Lambda(b_1
+b_2ln(\frac{\Lambda}{\mu}))\;.
\end{equation}
A diagram-by-diagram investigation shows that
all the diagrams that contribute to $b_2$
contain one of the diagrams in Fig.~11
as a sub-diagram.
This implies the four-fermi vertex
at order $G^2\alpha^2$ has
no primary logarithmic divergence.
Again, $b_1$ is regularization and renormalization
dependent.

\end{document}